\documentclass[showpacs,amsmath,amssymb,aps,showkeys,floatfix]{revtex4}
\usepackage{subfigure}
\usepackage{dcolumn}
\usepackage{bm}
\usepackage{epsfig}
\usepackage{amsfonts}
\usepackage{amssymb,amscd}
\def\lsim{\raise0.3ex\hbox{$<$\kern-0.75em\raise-1.1ex\hbox{$\sim$}}}
\def\gsim{\raise0.3ex\hbox{$>$\kern-0.75em\raise-1.1ex\hbox{$\sim$}}}

\newcommand{\be}{\begin{equation}}
\newcommand{\ee}{\end{equation}}
\newcommand{\ba}{\begin{eqnarray}}
\newcommand{\ea}{\end{eqnarray}}

\def\beq{\begin{equation}}
\def\eeq{\end{equation}}
\def\beqa{\begin{eqnarray}}
\def\eeqa{\end{eqnarray}}

\def\gappeq{\mathrel{\rlap {\raise.5ex\hbox{$>$}}
{\lower.5ex\hbox{$\sim$}}}}
\def\lappeq{\mathrel{\rlap{\raise.5ex\hbox{$<$}}
{\lower.5ex\hbox{$\sim$}}}}
\def\Toprel#1\over#2{\mathrel{\mathop{#2}\limits^{#1}}}

\begin{document}

\title{\Large Historical introduction to ultra peripheral collisions \footnote{Presented at  the international workshop on the physics of Ultra Peripheral Collisions (UPC2023), Playa del Carmen, Mexico, (October 2023).}}

\author{C.A. Bertulani}

\address{Department of Physics and Astronomy, Texas A\&M University-Commerce, Commerce, Texas, 75025, USA
}

\begin{abstract}
This is a brief history of photons, both soft and hard, real and virtual. About 150-100 years ago, Maxwell and Einstein discovered intriguing properties of electromagnetic fields and how to understand them both macroscopically and microscopically. Decades later, physicists developed the theory of renormalized quantum electrodynamics (QED), an incredibly accurate theory describing interactions of photons and other particles. Photons are used everywhere in academia and technological devices, from supermarket lasers and doors to academic studies in atomic, nuclear, and particle physics. In this article, I attempt to convey how the field of relativistic heavy ions rediscovered ultra-peripheral collisions (UPC) as a source of intense, almost real photons, and how it permits the study of a plethora of phenomena in the aforementioned academic fields. These phenomena are not always accessible by other means.
\end{abstract}



\maketitle

\section{Phriends and Photons}

German professors used to wield significant influence. Today some still maintain their power while most lament its decline. Prior to 2002, most academic university departments operated under a hierarchical structure, with a principal C4 professor at the apex, overseeing a pyramid of other C2 and C3 professors (following the hierarchy $C4 > C3 > C2$). This was the landscape into which I stepped upon my arrival at the University of Bonn and the KFA-J\"ulich laboratory in Germany in 1984. Upon my arrival, the C4 professor instructed me to approach the offices of three $C_<$ professors and inquire if they had a PhD project for me. So, I followed his directive. The first $C_<$ professor filled the blackboard with numerous Feynman-like diagrams and equations for complex many-particle systems. Daunted by the complexity, I was relieved when, after an hour, he explained that if we published three papers in Nuclear Physics A or Physical Review C within the next four years, I could write my thesis.

The scenario repeated itself in the second office, leaving me in awe once more as the expectation remained: three papers $\rightarrow$ thesis. Can you see the significance? Three or more published papers signified both the achievement and the prestige of attaining a Doctorate in Philosophy. It was in the third office that I met Professor Dr. Gerhard Baur, a $C_<$ professor. He candidly admitted that he didn't have an immediate plan for a PhD student but instead handed me a paper to peruse and discuss with him later. Finding the paper intriguing, he proposed that we reach deeper into its concepts, suggesting that we could expand upon its physics together. The topic seemed absurd to my fellow PhD student colleagues: heavy ion collisions without actual collisions, C'mom are you kidding me?. Yet, over the course of the next four years, Gerhard and I published 15 (fifteen) papers on the topic. This is equivalent to 5 PhD theses, according to the reigning J\"ulich standards at that time.

Back in the mid-80s, the C4 Professor avidly followed the works of a Japanese physicist named T. Suzuki, frequently publishing his works in the Physics Review Letters. Determined to bring him to J\"ulich for a six-month period, the professor invested significant time and effort, only to discover that there were three T. Suzukis, and he had invited the wrong one. By then, it was too late. Such mishaps were not uncommon during that era.
I vividly recall an encounter with the renowned T. Suzuki himself, who once stopped by my office and inquired about my PhD thesis. After my earnest attempt to convey the interest and significance of my research on ultra-peripheral heavy ion collisions (UPCs), he dismissed it as totally unimportant and undeserving of a conversation and left the room without saying goodbye.  

In another occasion, in 1986, during a workshop in Erice, Italy, Gerhard and I faced harsh criticism from nuclear physics experts who loudly rebuked our proposal regarding a double giant resonance in nuclei within UPCs. I can still recall my former colleague Eric Ormand likening the situation to ``opening a smelly can of worms."  In 1987, in a significant twist of fate, the late Gerry Brown, who served as the editor of Physics Reports at the time, expressed interest in publishing my PhD thesis~\cite{BeBa88}. This time I felt vindicated, knowing that Gerry was a visionary man.

Over the years, despite facing numerous embarrassing situations while attempting to promote UPCs as a viable source of useful physics, we persevered in our work. Particularly, Gerhard's dedication led to numerous contributions that extended beyond the 1990s. Gradually, UPCs emerged as a thriving field within both nuclear and particle physics.

I offer here a partial glimpse into the evolution of UPCs since the publication of my PhD thesis in 1988~\cite{BeBa88}. Despite the prevailing skepticism at the time (as Mark Strickman recalled during this meeting, ``in 1988, it looked like science fiction"), it's truly satisfying to witness how the predictions we put forth in works we published during the 1980s and 1990s have catalyzed a wealth of experimental findings and theoretical advancements. The domain of UPC physics has emerged as a focal point of extensive global research, playing a pivotal role in advancing our comprehension and consolidation of various phenomena within both QED and QCD realms \cite{KGS97,BHT02,BKN05,BHT07,Ba08,BeB94,Cont15,KN17,Nyst07,Klein17,KNS17,KS20,Stein21}.  

Please note that this account is not exhaustive, and I may inadvertently overlook many references. Whenever feasible, I will include a selection of equations utilized to make estimates or even conduct rigorous calculations for cross sections and probabilities in UPC processes.

\begin{figure}[t]
  \includegraphics[width=\columnwidth]{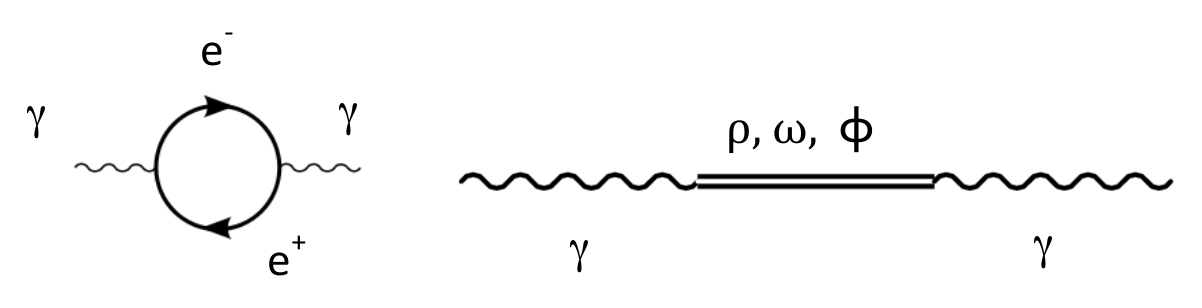}
  \caption{\label{banksy}The photon wavefunction contains ``hidden" quantum fluctuations into pair of particles, very much like a Banksy drawing.}
\end{figure}

\section{The Equivalent Photon Method}\label{subsec:prod}
As Gerhard and I digged deeper into the topic of UPC, we stumbled upon Enrico Fermi's practical method formulated back in 1924~\cite{Fermi:1924,Fermi:1925}. He explored atomic ionization induced by $\alpha$-particles. Fermi's work was published in German at the Zeitschrift f\"ur Physik and in Italian at the Nuovo Cimento. It's conceivable that Nuovo Cimento, established in 1923, initially functioned primarily as a repository of works by members of the Italian Physical Society (SIF) rather than adhering strictly to the norms of a traditional journal. In this context, Fermi might not have been influenced by the ``publish or perish" philosophy as much as we do.

My first collaboration with Gerhard resulted in the publication of our work in 1985~\cite{BeBa85}. Our objective was to extend Fermi's ``equivalent photon method" using quantum mechanics and first-order perturbation theory. We demonstrated that the electromagnetic fields of a highly energetic charge can induce  excitation processes in a nucleus and the matrix elements for the transition are equivalent to those induced by real photons. The final cross section can be expressed as a sum over multipoles, 
\begin{equation}
\sigma = \sum_{E/M,L}\int {d\omega  \over \omega} n_{E/M,L}(\omega) \sigma_\gamma^{(E/M,L)} (\omega), \label{epn2}
\end{equation} 
where $\sigma_\gamma^{(E/M,L)} (\omega)$ represents cross sections by (real) photons with energies $\omega$. Electric (E) and magnetic (M) ``multipolarities" include components of the photon angular momentum $L$. The  $n_{E/M,  L}(\omega)$ also depend on the beam energy $E_{\rm beam}$ and the photon energy $\omega$. They are denoted by ``equivalent (virtual) photon numbers" (EPN)~\cite{BeBa85}. For projectile bombarding energies below a few GeV/nucleon, the EPNs strongly depend on the $E/M,L$ multipolarity, e.g., $n_{E2} > n_{E1} > n_{M1}$, whereas at much larger energies they are approximately equal, $n_{E2} \sim n_{E1} \sim n_{M1}$, except for small excitation energies $\omega\ll \gamma/b$~\cite{BeBa85}. $\gamma = (1-v^2)^{-1/2}$ is the Lorentz factor,  $v$ is the projectile velocity, and $b$ signifies the collision the impact parameter.

The UPC community primarily focuses on the high-energy limit, where the following relation holds:
\begin{equation}
\sigma \simeq \sum_{E/M,L}\int {d\omega \over \omega} n(\omega) \sigma_\gamma^{(E/M,L)} (\omega) = \int {d\omega \over \omega} n(\omega) \sum_{E/M,L}\sigma_\gamma^{(E/M,L)} (\omega) = \int {d\omega \over \omega} n(\omega) \sigma_\gamma (\omega), \label{epn3}
\end{equation}
with $ \sigma_\gamma = \sum_{E/M,L}\sigma_\gamma^{(E/M,L)}$ representing the total cross section induced by a real photon. However, Eq. \ref{epn2} possesses a certain elegance as it delineates how heavy-ion collisions facilitate the differentiation of various photon multipolarities $E/M,L$. 

The prevalent belief that the EPN method, as described above for UPCs, is exclusively applicable to relativistic heavy ion collisions is erroneous. Equation \ref{epn2} remains valid whenever first-order perturbation theory is accurately enough to describe the UPC process under examination. This validity stems from the fact that in UPCs, because the charged particle does not penetrate the nucleus, conditions are such that $\nabla \times \mathbf{B} = 0$ and $\nabla \cdot \mathbf{E} = 0$, with $\mathbf{E}$ and $\mathbf{M}$ being the fields generated by the moving charge.

Another common fallacy is the notion that it is impossible to separate pure electromagnetic  processes from those involving the strong interaction. Various methodologies have been employed to differentiate between these processes including detailed studies of the cross sections dependencies on the projectile bombarding energy,  angular distributions, and on the nuclear excitation energies. Notably, employing light  nuclei (e.g., carbon) alongside heavy nuclei (e.g., lead)  in measurements aids in isolating the contributions of the two interactions.
\begin{figure}[t]
\begin{minipage}{17pc}
\includegraphics[width=17pc]{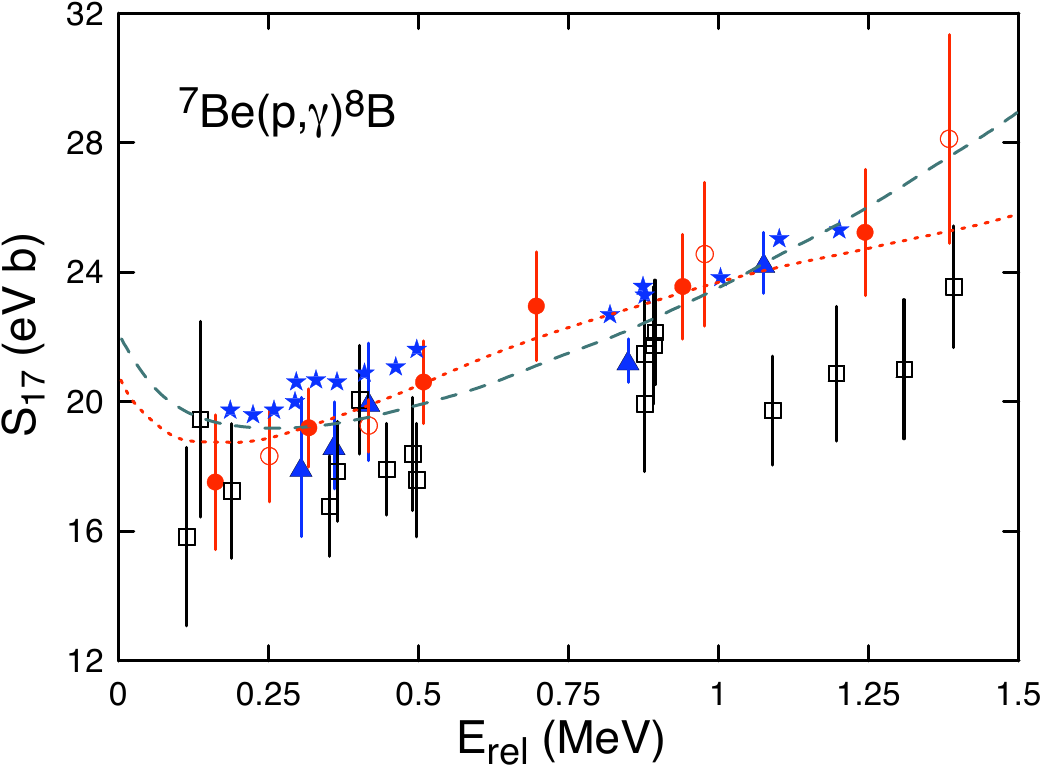}
\caption{\label{p7Be}Radiative capture reaction $^7$Be(p,$\gamma$)$^8$B. The dashed line is the no-core shell-model calculation of Ref.~\cite{Nav06} and the dotted line is from the resonant group method calculation of Ref.~\cite{Des94}.  Experimental data are from Refs.~\cite{Vau70,Fil83,Bab03,Jung03,Iwa99,Dav01,Schu03,Kaw69}.}
\end{minipage}\hspace{3pc}%
\begin{minipage}{17pc}
\includegraphics[width=17pc]{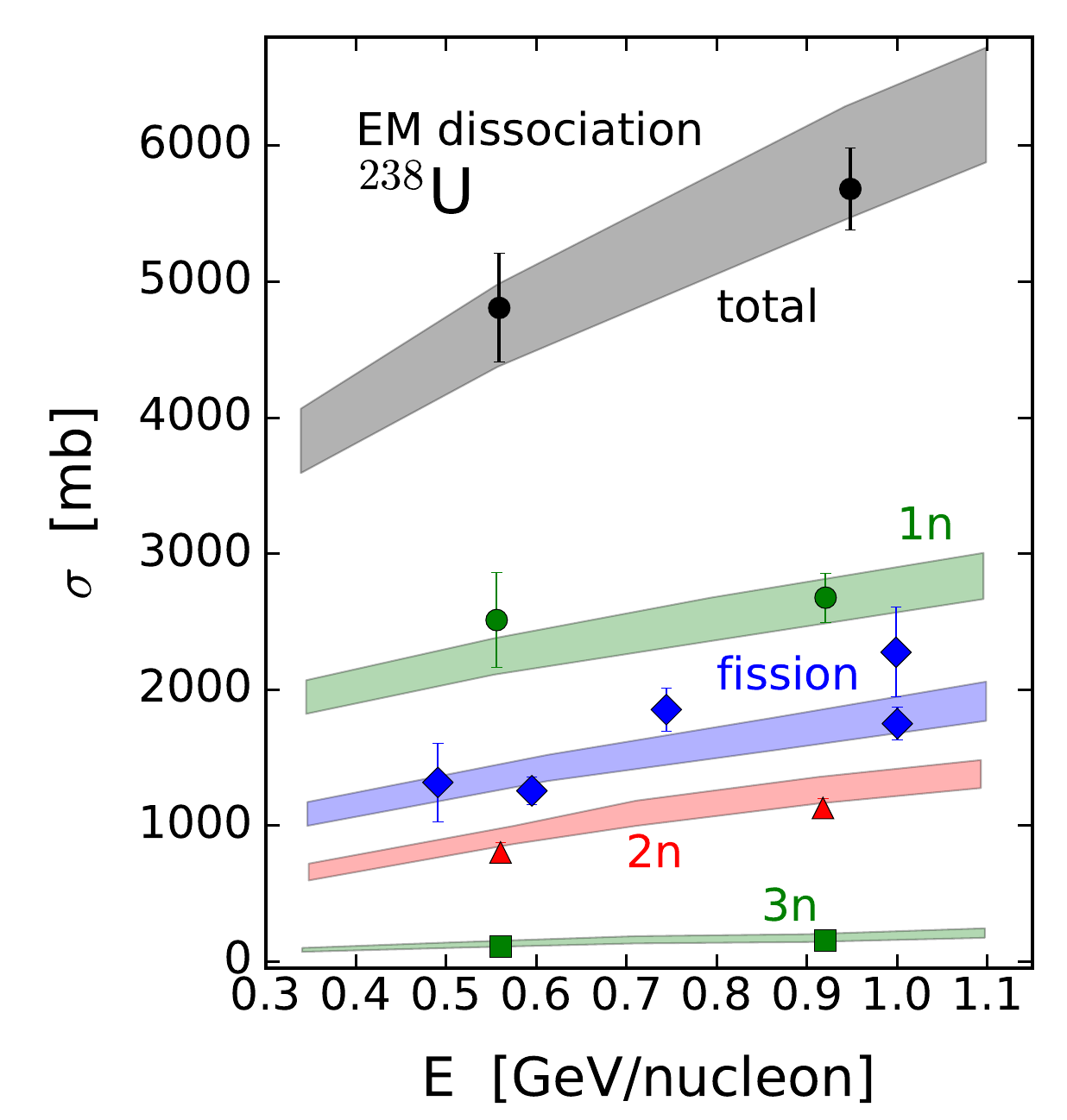}
\caption{\label{EMdis}Cross-section for EM dissociation of $^{238}$U. The excited nucleus decays by fissioning (diamonds) and  multiple neutron emission ($xn$)~\cite{Pol94,Arm96,Rub95,Aum96}.
 Theoretical calculations and their uncertainty bands are also shown~\cite{ABS95}.}
\end{minipage}
\end{figure}

However, there's a caveat. Equations \ref{epn2} and \ref{epn3} hold true only to first order. The photon, unlike electromagnetic waves with cute little fish-like patterns, resembles more of a ``Banksy" art piece~\cite{banksy}, an intricate octopus-like entity (see Figure \ref{banksy}). It exhibits fluctuations into other particles through emission and re-absorption processes, imprinting its wave function with their characteristics:
\begin{eqnarray}
\left| \gamma \right\rangle &=& C_{bare} \left| \gamma_{bare}\right\rangle + C_{ee} \left| e^{-}e^+ \right\rangle+\cdots + C_{qq} \left| q\bar{q} \right\rangle+
C_\omega \left| \omega\right\rangle + C_\phi \left| \phi\right\rangle +C_\rho \left| \rho\right\rangle +\cdots
\end{eqnarray}
Given the spin-parity $J^P = 1^{-}$, the photon can fluctuate into vector mesons ($\rho, \omega, \phi, J/\psi$) comprised of quark-antiquark pairs (following the vector dominance model). Furthermore, the photon often manifests as multiple energy components, some of which are invisible or virtual, akin to the enigmatic nature of Banksy's works~\cite{banksy}. Multistep processes involving photon splitting, emission of its ``parts," generation of intermediate states, or recycling are prevalent and integral to its nature. 

\begin{figure}[t]
\begin{minipage}{19pc}
\includegraphics[width=19pc]{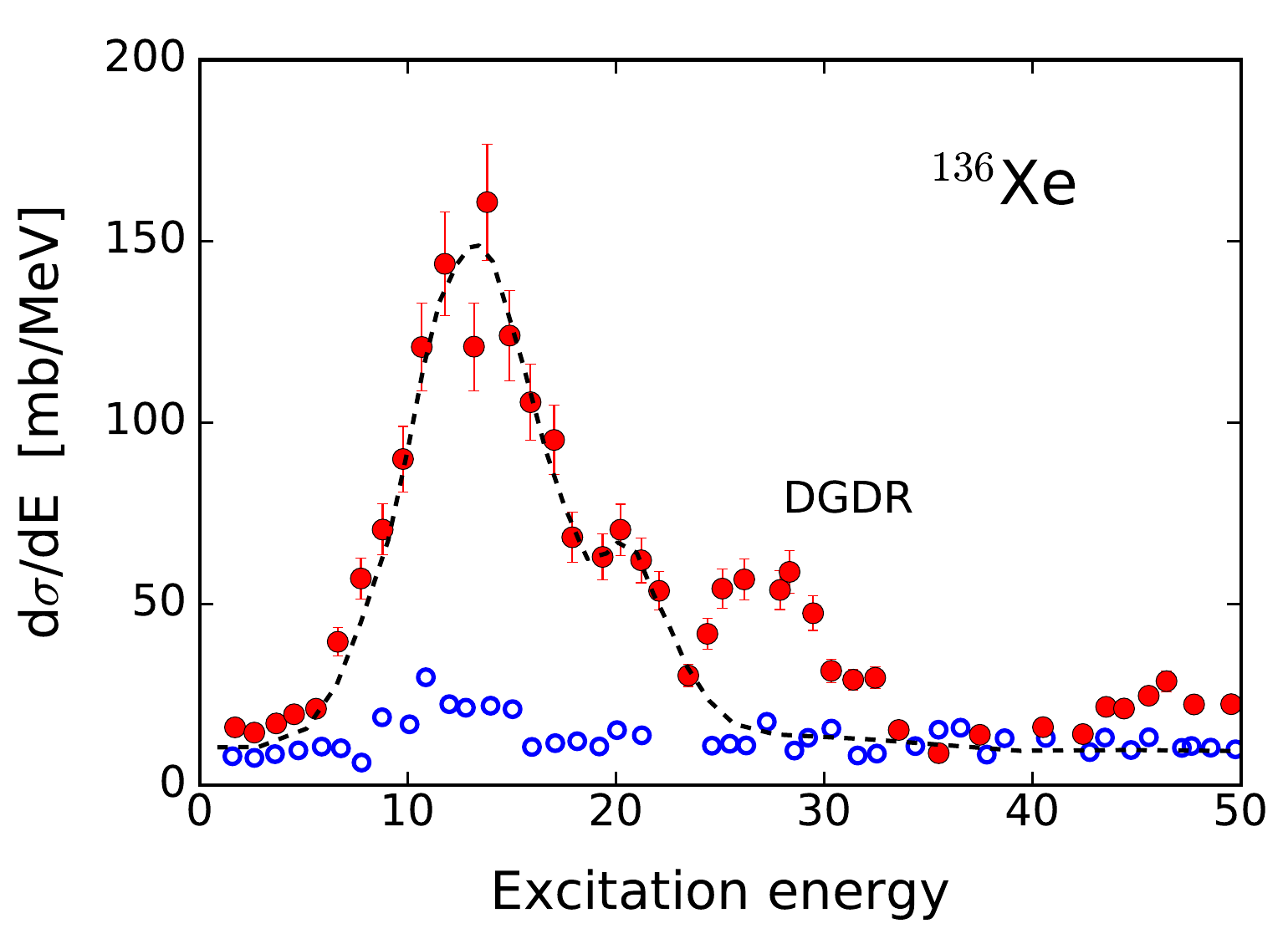}
\caption{Energy spectrum of giant resonances in $^{136}$Xe projectiles in UPCs with a large-Z target. The Giant Dipole Resonance (DGDR) is evident as a distinct peak~\cite{Sch93}.}
\label{fig:dgdr}
\end{minipage}\hspace{3pc}%
\begin{minipage}{16pc}
\includegraphics[width=16pc]{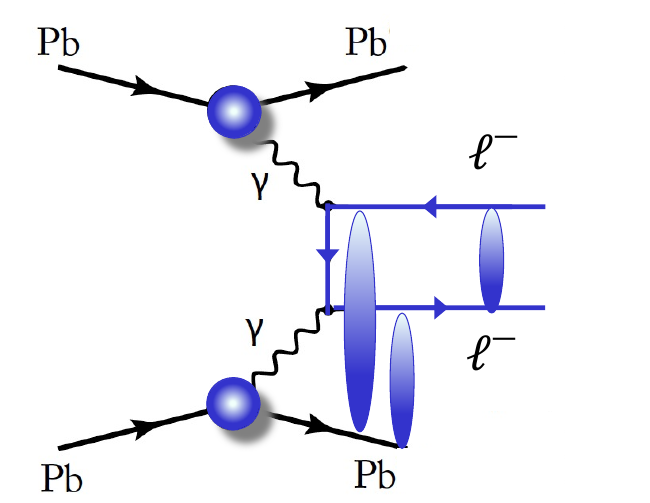}
\caption{\label{llpair}Production of lepton-antilepton in UPCs. Higher-order processes, including the production of multiple pairs, might be important.}
\end{minipage}
\end{figure}

\section{Applications in Low Energy Nuclear Physics}

\subsection{The Coulomb dissociation method}

In our proposal to employ UPC for nuclear physics, we introduced the Coulomb dissociation method in 1986~\cite{BBR86}. The differential Coulomb breakup cross section in UPC, where a projectile nucleus $a$ interacts with a target nucleus $A$ resulting in the reaction $a+A \longrightarrow b+c+A$, is expressed as:

\begin{equation}
\frac{d\sigma_{C}^{E/M,L}(\omega)}{d\omega d\Omega} = \frac{dn^{E/M,L}(\omega;\theta;\phi)}{d\omega d\Omega} \cdot \sigma_{\gamma+a \rightarrow b+c}^{E/M,L}(\omega).\label{CDmeth}
\end{equation}
Here, $\omega$ denotes the excitation energy provided by the relative motion, and $\sigma_{\gamma+a \rightarrow b+c}^{E/M,L}(\omega)$ represents the photonuclear cross section for the  photon energy $\omega$ and multipolarity ${E/M,L}$. The function $dn^{E/M,L}/d\omega d\Omega$ denotes the equivalent photon number, dependent on the  scattering angle $\Omega=(\theta,\phi)$~\cite{BBR86}.

Time reversal symmetry enables the deduction of the radiative capture cross section for the reaction  $b+c\longrightarrow a+\gamma$ from the experimentally obtained $\sigma_{\gamma+a \rightarrow b+c}^{E/M,L}(\omega)$. This methodology has proven invaluable in determining radiative capture cross sections for various reactions pertinent to astrophysics. An illustrative example is the $^{7}$Be($p,\gamma)^{8}$B reaction, initially investigated in Ref.\cite{Tohru}, followed by numerous subsequent experiments. This is shown in Figure \ref{p7Be}, where the red dots were obtained using the Coulomb dissociation method and  the experimental data are from Refs.~\cite{Vau70,Fil83,Bab03,Jung03,Iwa99,Dav01,Schu03,Kaw69}. Further discussions on the outcomes derived through this method can be found in Refs.\cite{EBS05,Mosh06,Adel11}.

Equation \ref{CDmeth} is rooted in first-order perturbation theory and assumes that the nuclear contribution to the breakup is either negligible or separable under specific experimental conditions. The influence of nuclear breakup has been scrutinized by several researchers (see, e.g., Ref.~\cite{Ber23}). Weakly-bound nuclei, such as ``halo nuclei," characterized by very small neutron separation energies, exhibit significant multiple-step or higher-order effects, particularly through continuum-continuum transitions, as demonstrated in Ref.~\cite{Ber23}.

\subsection{Giant Resonances} 

Another application of UPCs in low-energy nuclear physics was in investigating giant resonances within nuclei~\cite{BeB86}. Typically, it decays by neutron emission, and for energies around 1 GeV/nucleon, such as those achievable at the GSI laboratory in Germany, the excitation cross sections can reach several barns~\cite{ABS95,ASBK96} (see Figure \ref{EMdis}). At these energies, giant resonances can also be efficiently excited through nuclear interactions. However, it was soon recognized that for nuclei with high atomic numbers (large-Z), cross sections are significantly smaller compared to those induced by electromagnetic (EM) interactions~\cite{BeBa88,Pol94,Arm96,Rub95,Aum96}. Presently, Coulomb excitation and decay of giant resonances serve as a valuable experimental tool, including in studies related to nuclear fission~\cite{Aum97,BKL20,Stet15}. Because of its substantial cross section, this process has been proposed for use as a heavy-ion collider luminosity monitor~\cite{BeB94,BCW98,Klein01}.

 \subsection{Multiphonon giant resonances} 
  \begin{figure}[t]
\begin{minipage}{20pc}
\includegraphics[width=20pc]{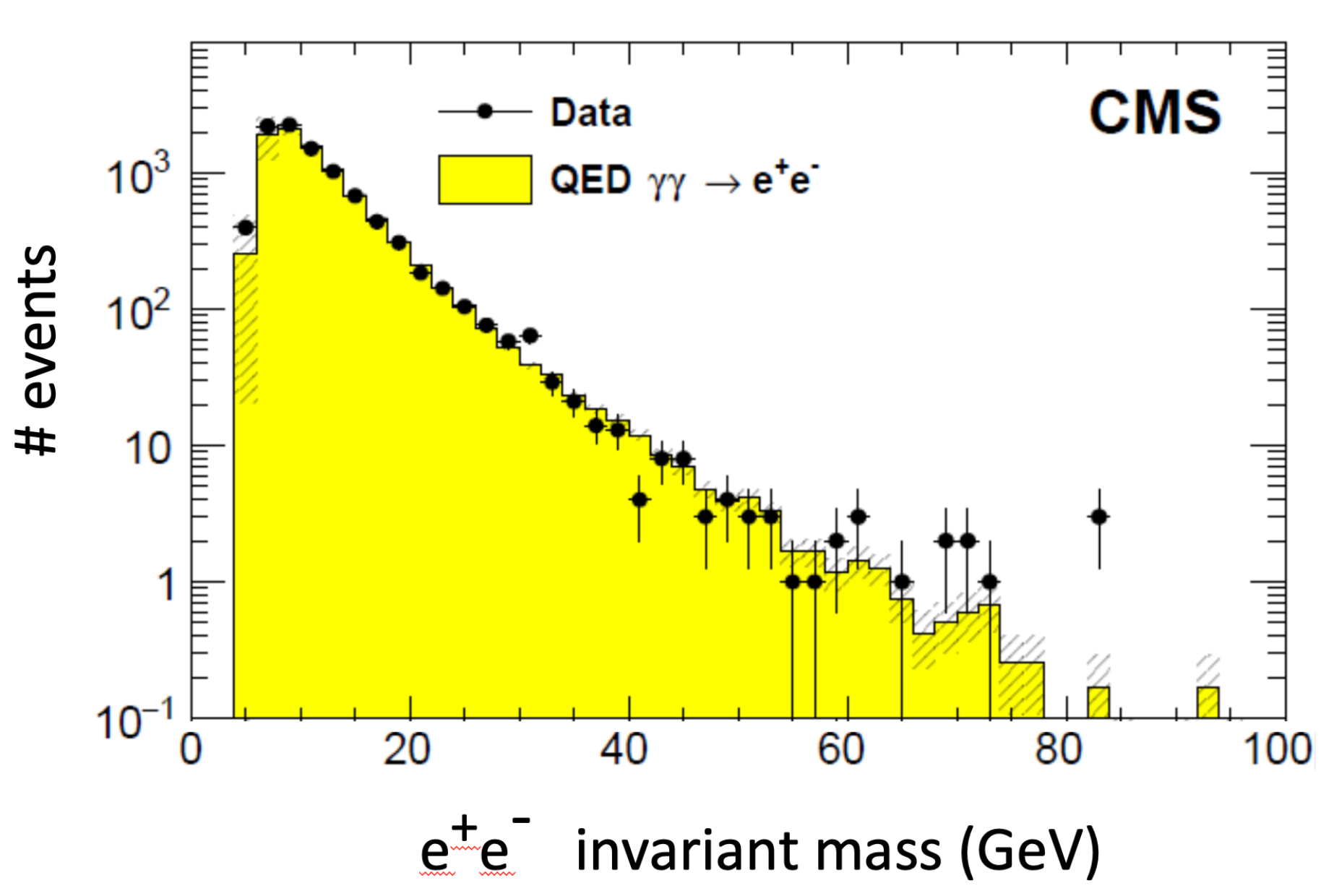}
\caption{\label{cms9}$e^+e^-$ pairs produced in UPCs observed with the CMS detector (from Ref.~\cite{Ada18}).}
\end{minipage} \hspace{2pc}
\begin{minipage}{15pc}
\includegraphics[width=15pc]{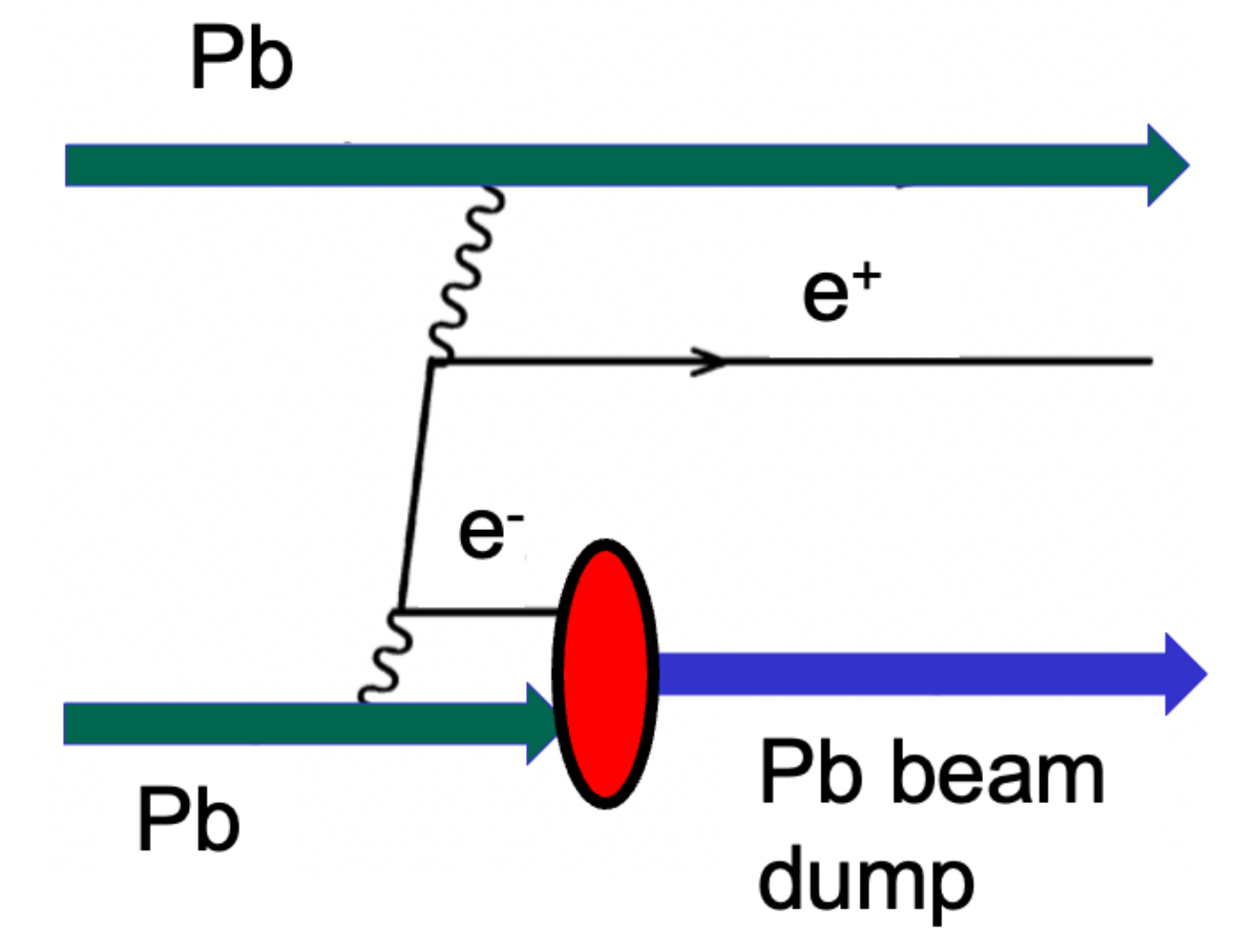}
\caption{\label{llpairc}Pair production with electron capture leading to possible beam losses \cite{BeB94}.}
\end{minipage}\hspace{3pc}%
\end{figure}
 
In 1986, Gerhard and I proposed the excitation of multiple giant resonances in UPCs through multiple photon exchange~\cite{BeBa86,BeBau86}. The non-perturbative treatment of this process can be achieved using coupled-channels equations. Additionally, Glauber methods offer a means to account for diffraction effects arising from nuclear interactions. Predictions indicated significantly large excitation probabilities for double, triple, and multi-phonon resonances in nuclei~\cite{BeBa86,BeBau86}. Two pioneering experiments conducted at the GSI laboratory in Germany in 1993 provided empirical support for these predictions~\cite{Sch93,Rit93}. One experiment employed gamma-gamma coincidences to identify the decay of the double giant dipole resonance (DGDR)~\cite{Rit93}, while the other observed predominantly neutron emission as a result of the decay of the multi-phonon giant resonances~\cite{Sch93} (see Figure \ref{fig:dgdr}). The study of DGDR is particularly noteworthy as its strength and width offer valuable constraints for nuclear models regarding the absorption of multiple photons. Comprehensive reviews on this topic are available in references~\cite{Em94,ABE98,BP99}.

\section{Pair production} 

One of the processes that called our attention in 1986 was the abundant production of electron-positron pairs in UPCs (Figure \ref{llpair}). The pioneering studies on the production of $e^+e^-$ pairs in UPCs trace back to the 1930s. Bethe, Racah, Bhabha, Tomonaga, Nishina, Furry, among others, developed techniques to compute pair production employing the newly formulated Dirac equation. Dirac's equation famously predicted the existence of the positron, conceptualized as a void in the ``vacuum sea" of electrons. Initially, the exploration for this ``void" (positron) was pursued via UPCs involving cosmic rays possessing high kinetic energies $E$~\cite{FC33,Bet34,Bha35,NTT34,NTK35,Rac37}.

Under the assumption that the energy of produced pairs vastly exceeds the electron rest mass $m_e$, almost all theoretical forecasts~\cite{FC33,Bet34,Bha35,NTT34,NTK35,Rac37} yielded a production cross section equal to
\begin{equation}
\sigma_{e^+e^-} ={28\over 27 \pi}  \left(Z_1Z_2 \alpha r_e\right)^2 \ln^3 \left({\gamma\over 2}\right), 
\end{equation}
where the Lorentz factor is $\gamma \simeq E/M$, with $M$ the mass of the cosmic ray ion, and $r_e=e^2/m_e$ fm is the classical electron radius (for a concise discussion, see Ref.~\cite{BeBa88}). The cross section reaches scandalous values of 200 kbarn for PbPb at LHC. In 1986, we revisited these computations employing contemporary methods in Quantum Electrodynamics (QED), beyond the scope of the physics known in the 1930s. We developed a theory incorporating final state interactions (depicted as blobs in Figure \ref{llpair}) utilizing Bethe-Maximon distorted waves \cite{BeBa88}. Given the significantly large cross sections, on the order of kilobarns for standard collider energies like those at the LHC, we demonstrated the pertinence of exploring higher-order corrections \cite{BeBa88,BaBe89}. 

The simplest approximation we devised for multiple pair production assumed a Poisson distribution for the production probability for $n$ pairs at a given impact parameter so that 
\begin{equation}
P_{e^+e^-}(b) = {[P_0(b)]^n \over n!} \exp\left[-P_0(b)\right],
\end{equation} 
where $P_0(b)$ is the probability for a single pair production, which can be calculated analytically~\cite{BeBa88}, 
\begin{equation}
P_0(b) = {14 \over 9 \pi^2 b^2} \left(Z_1Z_2 \alpha r_e\right)^2 \ln^2 \left( \gamma\over 2 m_e b\right),
\end{equation} 
valid for $\gamma/m_e >b>1/m_e$~\cite{BeBa88}. The effect of final state interactions in multiple pair production was further investigated in numerous other publications and reviews~\cite{Henck95,Baur02,BHT07,Ba08,Ba08b}.
\begin{figure}[t]
\begin{minipage}{17pc}
\includegraphics[width=17pc]{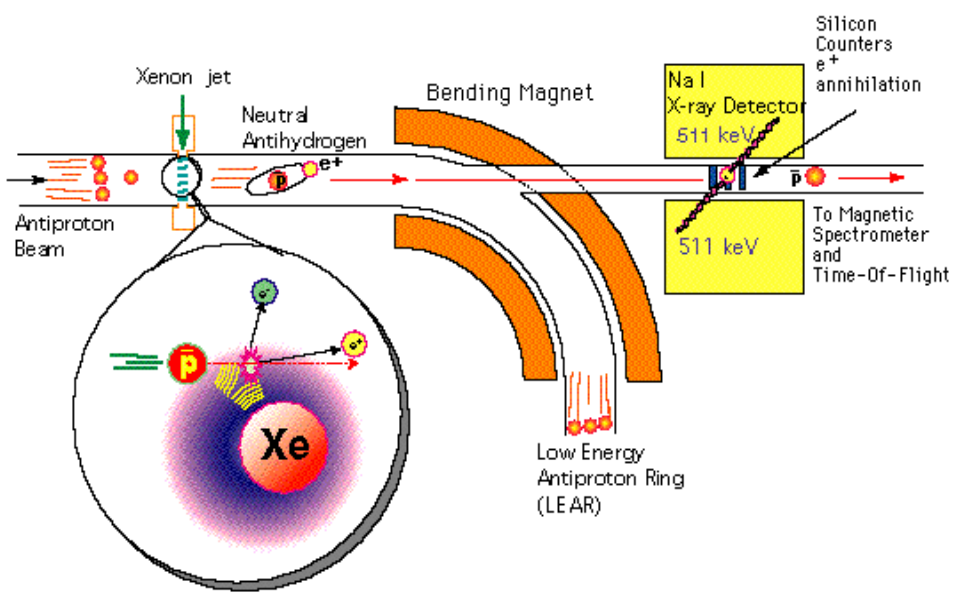}
\caption{\label{setup}Experimental setup employed in the identification of anti-atoms at the LEAR/CERN in 1996.}
\end{minipage} \hspace{2pc}
\begin{minipage}{18pc}
\includegraphics
[width=18pc]
{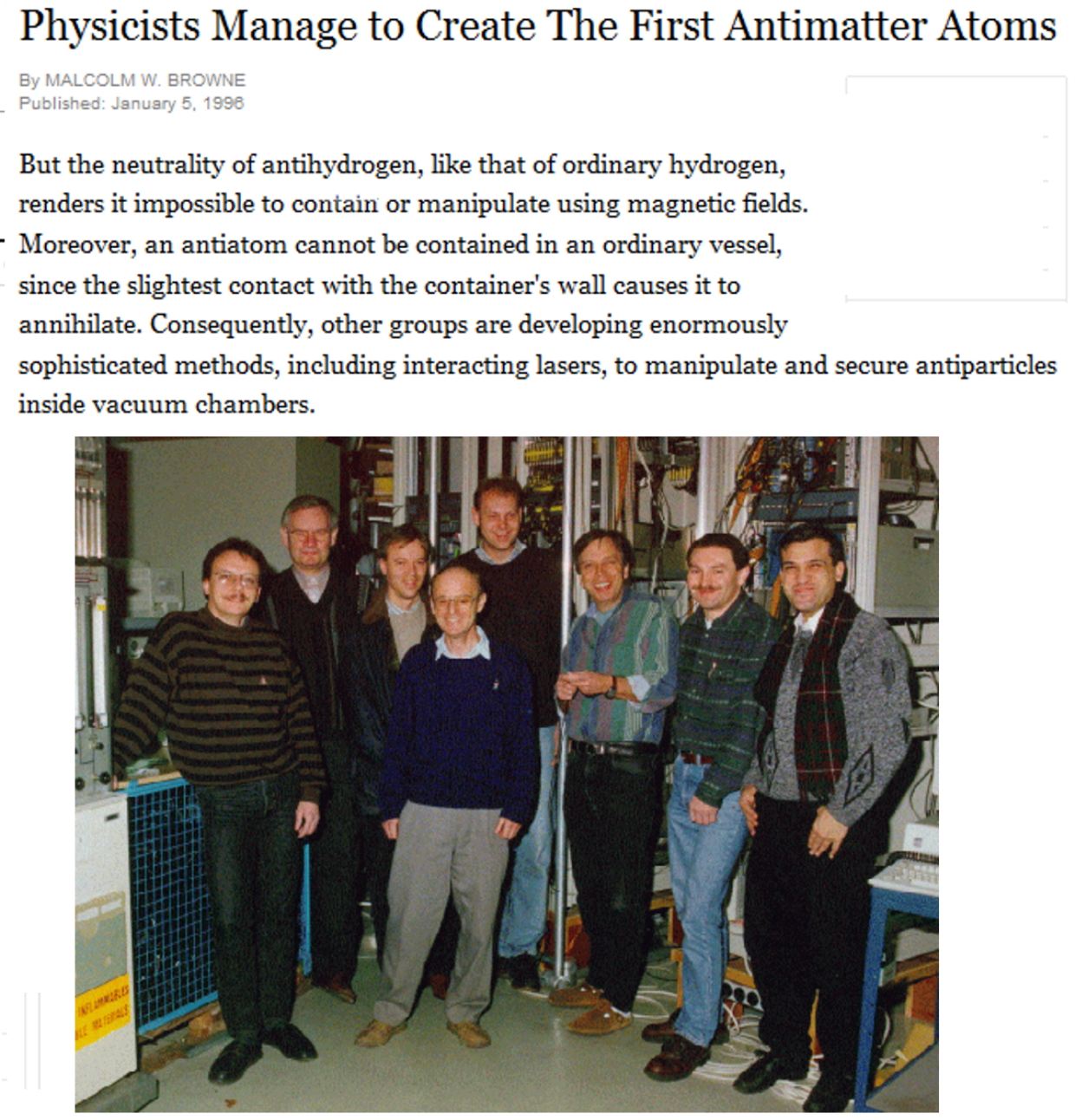}
\caption{\label{NYTimes}The New York Times report on the first ever production of an anti-atom in the laboratory (1996). Gerhard Baur (a theorist) stands in the center of the picture.}
\end{minipage}
\end{figure}

It is gratifying to witness current LHC experiments measuring the foundational pair-production cross sections initially calculated in the 1930s and later studied in References~\cite{BeBa88,BaBe89,Henck95,Baur02}. These processes are now largely comprehended theoretically. Initial experiments validated predictions grounded in QED with the STAR detector at the Relativistic Heavy Ion Collider (RHIC)~\cite{Ada04} and later that with the CMS detector at CERN (Figure \ref{cms9}). The generation of other particle-antiparticle pairs, such as  (Figure 1), $\gamma \gamma\rightarrow \mu^{+}\mu^{-}$, $\gamma \gamma\rightarrow \pi^{+}\pi^{-}$, $\gamma \gamma\rightarrow W^{+}W^{-}$, etc.,  is not insignificant, as calculations in Refs.~\cite{BaBe87,Bau88,BaBe89} have shown, and confirmed in subsequents works~\cite{BaBe87,BeBa88,Bau88,BeB94}. For the production of $ \mu^{+}\mu^{-}$ ($\pi+\pi^-$) and $ \tau^{+}\tau^{-}$ pairs we can use the equations above if   $\gamma \gg 16$ for muon-pair production, or $\gamma \gg 200$ for tau-pair production. If these conditions are not satisfied, significant corrections to these equations are necessary, as shown in Ref.~\cite{BeBa88}. At the LHC, the Lorentz boost factor $\gamma$ in the laboratory frame is about 7000 for p-p, 3000 for Pb-Pb collisions and the conditions above apply.

\section{Anti-atoms and Exotic Atoms} 

Arguably, one of the most aesthetically captivating applications of UPCs  involves the generation of lepton pair production, coupled with the capture of the negative lepton by one of the colliding ions ~\cite{BeBa88,BeBa98,BeEl10} (see Figure \ref{llpairc}). The cross section for electron-pair production with capture of the electron in an atomic K-shell orbit is~\cite{BeBa88} 
\begin{equation}
\sigma_{K} ={33\over 20}  Z_1^5Z_2^5 \alpha^5 r_e^2 \ln \left({\gamma\over 2}\right). 
\end{equation}
A factor $\sum_n 1/n^3 \simeq 1.202$ increases the value of this cross section when capture is considered for all other orbits. However, the effects of electron screening and distortions in heavy atoms modify the electron capture cross sections appreciably~\cite{BeBa88}.     

A groundbreaking adaptation of this method was developed for anti-hydrogen production at the LHC, initially proposed in Ref.~\cite{MB94}, and later substantiated by a pioneering experiment conducted at CERN and documented in 1996 \cite{Bau96} (see Figure \ref{setup}). Operating within the Low Energy Antiproton Ring (LEAR), CERN's venture saw antiprotons colliding with protons, resulting in the capture of positrons into orbit around the antiproton. This landmark achievement marked the first terrestrial production of anti-atoms, with the detection of 11 anti-hydrogen atoms, attracting widespread attention from global media outlets including the New York Times~\cite{NYT96} (see Figure \ref{NYTimes}). Gerhard used to say that it is not difficult for scientists to publish in Nature, Science, or Physical Review Letters, but it is very difficult to get their picture and achievements appearing in the New York Times. ``I must have reached the apex of my career", he said.

Subsequently, a similar endeavor was undertaken at FERMILAB \cite{Bla98}, yielding 57 identified events, aligning with perturbative calculations performed prior to the experiment \cite{BeBa98} (see also Ref.~\cite{BD01}). Expanding horizons, investigations into the properties of anti-atoms are presently conducted utilizing ion traps, aimed at scrutinizing fundamental symmetries~\cite{Eug10,Gro10}. Furthermore, the ambit of UPCs extends to the production of larger antimatter entities such as anti-deuterium, anti-tritium, and anti-helium \cite{Aga11}. Predictions made in Ref.~\cite{BeEl10} extend the potential of UPCs, envisioning the production of muonic, pionic, and other exotic atoms through coherent photon exchange between ions at the LHC.

It's noteworthy that as early as 1988, in Ref.~\cite{Bau88}, the process of electron-positron production, with subsequent capture of the electron into an atomic orbit of one of the ions, was proposed as a source of beam loss in relativistic colliders. Initial estimations indicated potential degradation of a high-Z ion beam within 2 hours at the LHC. Subsequent theoretical scrutiny \cite{Klein14} and recent experimental validation at the LHC \cite{Schau20} corroborate the earlier expectations \cite{Bau88}. Another intriguing facet lies in the production of ortho- and para-positronium in UPCs. Ref.~\cite{BN02} furnishes a theoretical framework within quantum field theory to compute positronium and other bound-states, such as mesons (bound $q\bar q$), through $\gamma\gamma$- and $\gamma\gamma\gamma$-fusion in UPCs \cite{BBGN16,BGMN17,Far23}. 

One expects the existence of six leptonic atoms: (a) the positronium (e$^+ e^-$), (b) the muonium ($\mu^+ e^-$ ), (c) dimuonium ($\mu^+ \mu^-$ ), (d)
tauonium ($\tau^+ e^-$), (e) tau-muonium ($\tau+ \mu^-$), and (f) ditauonium ($\tau^+ \tau^-$). Only positronium, muonium, and dipositron-positronium, and (e$^+ e^-$)(e$^+ e^-$) were observed experimentally~\cite{Deu53,Hug60,Cas07}. 
Dimuonium has a radius a few hundred times smaller than the positronium and muonium. Because of its large mass, it is sensitive to physics beyond the standard model (BSM)~\cite{Fox22} and unexplored time-like regions of quantum electrodynamics (QED). Therefore, the discovery of the dimuonium would be a significant one in physics~\cite{Hug71}. The production of the dimuonium in UPC has been calculated and shown to be feasible to be measured at the LHC~\cite{Ginz98,Krac18,Azev20,Franc22}.  It has also been shown that the observation of the ditauonium at the LHC is possible~\cite{Dent23}. In both cases, one expects the identification to proceed via its displaced vertex with a rather good control of the combinatorial dimuon background.
\begin{figure}[t]
\begin{minipage}{15pc}
\includegraphics[width=15pc]{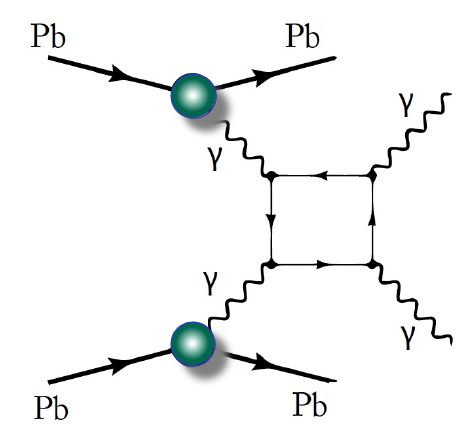}
\caption{\label{lbl}Light-by-light scattering using UPCs.}
\end{minipage}\hspace{3pc}%
\begin{minipage}{18pc}
\includegraphics[width=18pc]{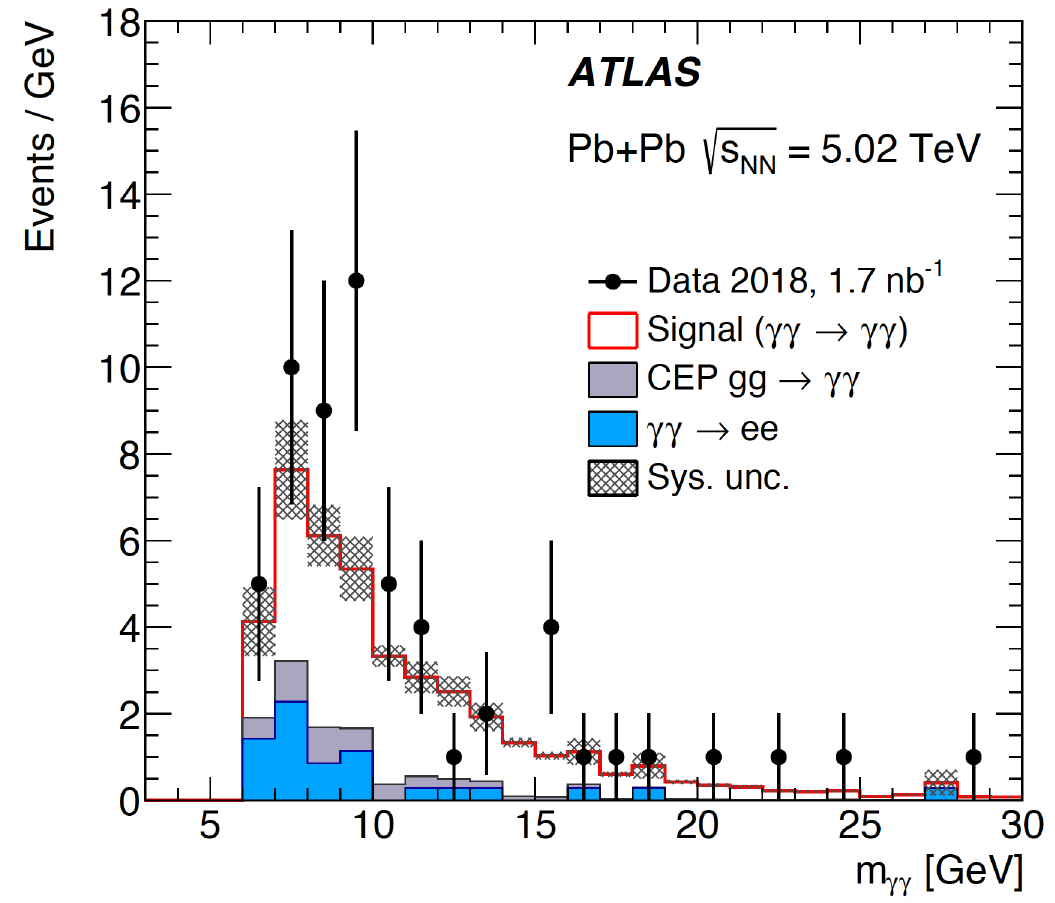}
\caption{\label{lblexp}Number of light-by-light scattering events observed with the ATLAS detector \cite{Aab17}.}
\end{minipage}
\end{figure}

\section{Light-by-light Scattering}

The phenomenon of elastic scattering of light by light, $\gamma + \gamma \rightarrow \gamma + \gamma$, exclusively occurs through the fluctuation of a photon into an particle-antiparticle pair (Figure \ref{lbl}). This process bears a relatively minuscule probability, rendering a direct study with real photons unattainable. In our pioneering work~\cite{Bau88}, Gerhard and I proposed the utilization of UPCs to probe $Z_1+Z_2 \rightarrow Z_1+Z_2 + \gamma+\gamma$, whereby two virtual photons scatter via a box diagram, producing two real photons. We underscored the theoretical uncertainty inherent in calculations reliant on the Delbr\"uck scattering formalism~\cite{Bau88}. To leading order, the cross section for $\gamma^*\gamma^*\rightarrow \gamma\gamma$ scattering events in UPCs leading to high energy photons is~\cite{Bau88}
\begin{equation}
    \sigma_D \sim 2.54 \times 10^{-2} Z^4\alpha^4 r_e^2  \ln^3 \left( { \gamma \over m_e R}\right) ,
\end{equation}
where  $R$ is the radius of the colliding ions. For PbPb collisions at the LHC, one gets huge cross sections about $\sigma_D\sim 50$ b. But not all these photons can be detected as purely due to this process, due to background by similar processes such as Bremsstrahlung.  For $E_\gamma > m_\mu$ and Pb + Pb collisions at the LHC, the cross section is much smaller, of the order of $\sigma_D\sim 30$ nb~\cite{BBN24}.

Light-by-light scattering was further studied in numerous other works where additional mechanisms for photon-photon scattering such as VDM-Regge, two-gluon exchanges, and meson resonances were considered (see, e.g., Refs.~\cite{Dent13,Mari16,Leb17,Klus19,Juc24}).  The ATLAS collaboration at the LHC achieved a breakthrough by observing this process experimentally (Figure \ref{lblexp})~\cite{Aab17}.  Such a revelation not only validates theoretical conjectures but also paves the path for exploring physics beyond the Standard Model (SM). An observed cross-section exceeding that predicted by the SM model could potentially signify the existence of new particles, such as axions~\cite{GS20,GMR21,Scho21}. This remarkable observation instigates a quest into realms of physics yet unexplored~\cite{Scho21}.

\section{Meson production in UPCs} 

Gerhard and I devised a simple way to extend the EPN method to compute the production of a bound-particle denoted as $X$ in UPCs (Figure \ref{jpsi})~\cite{Bau88}. The cross-section for photon-photon fusion can be expressed as 
\begin{equation}
\sigma_{X}=\int {d\omega_{1}\over \omega_1}{d\omega_{2}\over \omega_2} n_{\gamma}\left( \omega_{1}\right) n_{\gamma}\left(
\omega_{2}\right) \sigma^{X}_{\gamma\gamma}\left( \omega_{1}\omega_{2}\right),
\label{eq:two-photon}%
\end{equation}
where $n_{\gamma}(\omega)$ represents the EPNs for photon energies $\omega$, and $\sigma
^{X}_{\gamma\gamma}\left( \omega_{1}\omega_{2}\right) $ stands for the photon-photon cross-section for producing particle $X$. It can be computed using Low's expression \cite{Low60}, derived from the detailed balance theorem, connecting the $\gamma\gamma$-production with the $\gamma\gamma$-decay, 
\begin{equation}
\sigma^{X}_{\gamma\gamma}\left( \omega_{1}\omega_{2}\right) =8\pi^{2}(2J+1){\frac
{\Gamma_{m_{X}\rightarrow\gamma\gamma}}{m_{X}}}\ \delta\left( \omega_{1}\omega_{2}
-m^{2}_{X}\right) .\label{Low}%
\end{equation}
Here, $m_{X}$, $J$,  and $\Gamma_{m_{X}\rightarrow\gamma\gamma}$ denote the  mass, spin, and photon-photon $\gamma\gamma$ decay width of particle $X$, respectively. The delta-function ensures energy conservation. In Ref.~\cite{Bert09}, the significance of various meson models and exotic states is comprehensively discussed, encompassing states that were previously overlooked.

To segue, it's worth mentioning a proposal from 1989 aimed at detecting the Higgs particle using UPCs at the LHC \cite{Pap89}. This proposal elucidates the production mechanism through equations (\ref{eq:two-photon},\ref{Low}) while making appropriate assumptions regarding the Higgs properties. Our initial estimations in 1988 yielded a cross-section of 1 nb \cite{Bau88}, closely aligning with the  Higgs production cross sections at the LHC via hadronic processes, albeit with the advantage of minimizing the production of a multitude of other particles. However, it later became evident that the direct photon-photon production of $b\bar b$ pairs is significantly larger~\cite{Dree89}, although more optimistic scenarios for the Higgs production in UPCS have emerged later~\cite{DEnt10,DEnt20}. Since the primary mechanism for Higgs decay involves $b\bar b$ pairs, it can be inferred that  the Higgs production in UPCs would be overshadowed by a background of directly produced pairs. The elusive Higgs was eventually observed at the LHC in hadronic interactions~\cite{ATL12,CMS12}.

\section{Production of Exotic Mesons in UPCs}
Multiquark states, including multiquark molecules such as $(q\overline{q})(q\overline{q})$, glueballs $(gg)$, and hybrid mesons $(q\overline{q}g)$, hold significant role in meson spectroscopy \cite{Yao06}. UPCs offer a potential avenue for probing multiquark resonances by means of anomalous $\gamma\gamma$ couplings and multiquark energy spectra. They could serve as a means to test predictions concerning  ``abnormal" states \cite{BN02,BBGN16,BGMN17}. The $\gamma\gamma$ width serves as a gauge of the charge of constituent quarks, facilitating differentiation between quark resonances and gluon-dominated resonances (termed ``glueballs"). The failure to verify meson production via $\gamma\gamma$ fusion also serves as a significant indicator for the search for glueballs \cite{BN02,BBGN16,BGMN17}. In UPCs, a glueball is formed by the annihilation of a $q\overline{q}$ pair into gluon pairs. In contrast, normal $q\overline{q}$ mesons are created directly.

\section{Probing Parton Distribution Functions with UPCs} 

The exploration of vector meson production, like $J/\psi$ and $\Upsilon(1s)$, can be pursued utilizing Eq. (\ref{epn2}) and first calculations and experiments on vector meson production we published in Ref.~\cite{KN99,KN00,BKN02,Adl02,Abel08,PHE09}. In 2001, we  proposed~ \cite{Goncalves:2001vs} to use vector meson production for constraining generalized partonic distributions in nuclei, denoted as $F_A(x,Q^2)$, corresponding to a momentum fraction $x$, based on a formalism developed in Ref.~\cite{Frank98}. For the mechanism induced by real photons, we employed the relation
\begin{equation}
\left. {d\sigma^{\gamma A \rightarrow VA}\over dt }\right |{t=0} = {16 \pi \alpha_s^2 (Q^2) \Gamma_{ee}\over 3 \alpha M_V^5} \left[ xF_A(x,Q^2)\right]^2,
\end{equation}
where $\alpha_s(Q^2)$ represents the strong interaction coupling evaluated at the perturbative Quantum Chromodynamics (pQCD) factorization scale $Q^2 = W_{\gamma g} ^2$, $M_V$ denotes the vector meson mass, $\Gamma_{ee}$ signifies its leptonic decay width, and $x=M_V^2/W^2_{\gamma p}$ denotes the fraction of nucleon momentum carried by gluons.

\begin{figure}[t]
\begin{minipage}{17pc}
\includegraphics[width=17pc]{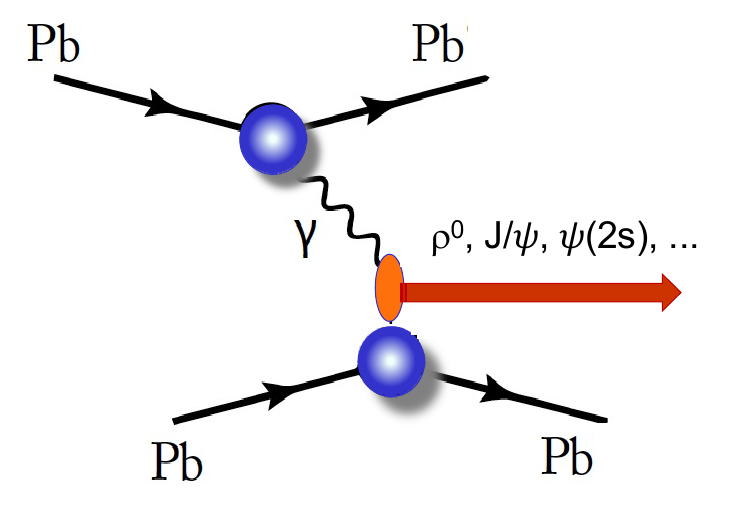}
\caption{\label{jpsi}Production of vector mesons in UPCs.}
\end{minipage} \hspace{3pc}%
\begin{minipage}{17pc}
\includegraphics[width=17pc]{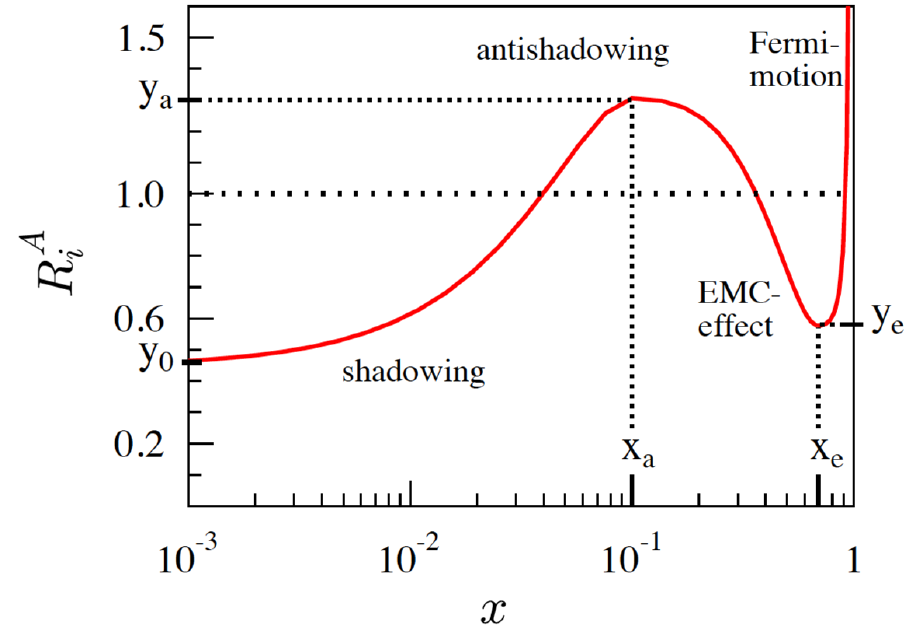}
\caption{\label{emc}Medium modification function displaying various effects as the momentum fraction $x$ varies.}
\end{minipage}
\end{figure}

The nuclear Parton Distribution Function (PDF), denoted as $F_{a}^A({\bf r},x,Q^2)$, can be expressed as a folding of a medium modification function $R_{a}^A({\bf r},x,Q^2)$ with  a nucleon PDF, represented as $f_{a}(x,Q^2)$. Here, the subscript $a$ signifies a parton species, while the superscript $A$ denotes a specific nucleus \cite{Adeluyi:2011rt,Adeluyi:2012ph}. The variable ${\bf r}$ denotes the nucleon coordinate within the nucleus.

Nuclear modifications are encapsulated within $R_{a}^A(x,Q^2)$. For values of $x$ less than 0.04, experimental observations reveal a shadowing effect, characterized by nuclear PDFs being smaller than the  free nucleon distributions, denoted as $R_a^A < 1$ (Figure \ref{emc}). In the range of $0.04 < x < 0.3$, an antishadowing effect is evident, with $R_a^A > 1$. The EMC effect manifests in the domain of $0.3 < x < 0.8$. Additionally, for $x > 0.8$, an enhancement also occurs  attributed to the nucleonic Fermi motion. These effects are governed by distinct underlying physical principles.

\begin{figure}[t]
\begin{minipage}{17pc}
\includegraphics[width=16pc]{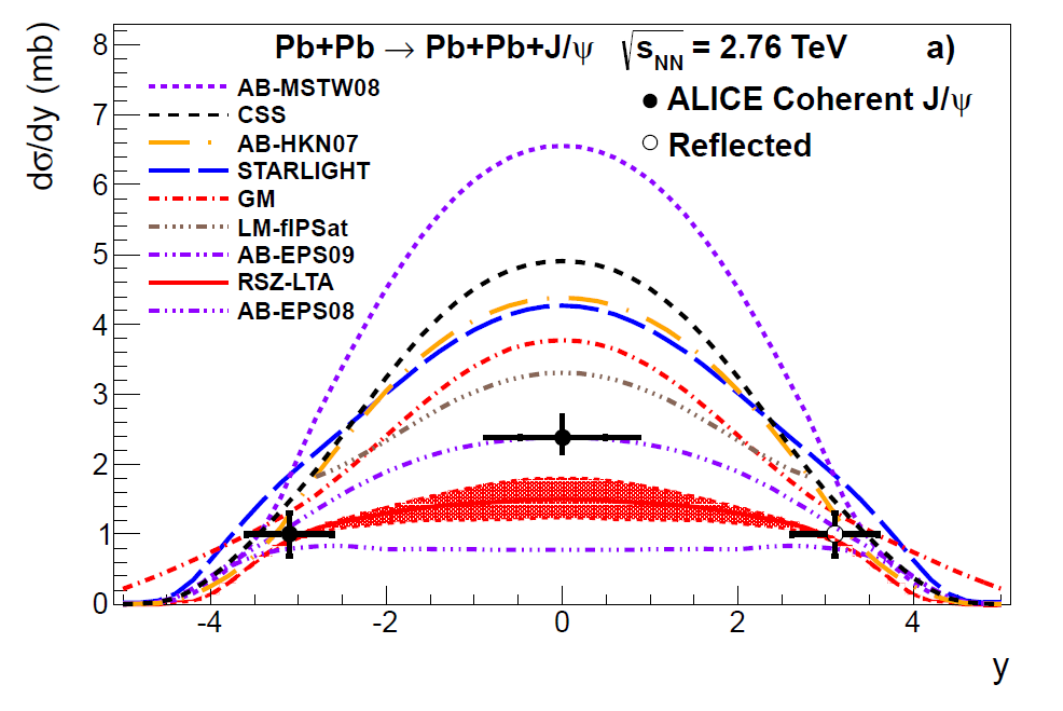}
\caption{\label{pdf1}Production of J/$\Psi$ in UPCs at the LHC as a function of the rapidity $y$. It serves as a probe of different PDFs (adapted from Ref.~\cite{Abe13}).}
\end{minipage}\hspace{2pc}%
\begin{minipage}{17pc}
\includegraphics[width=17pc]{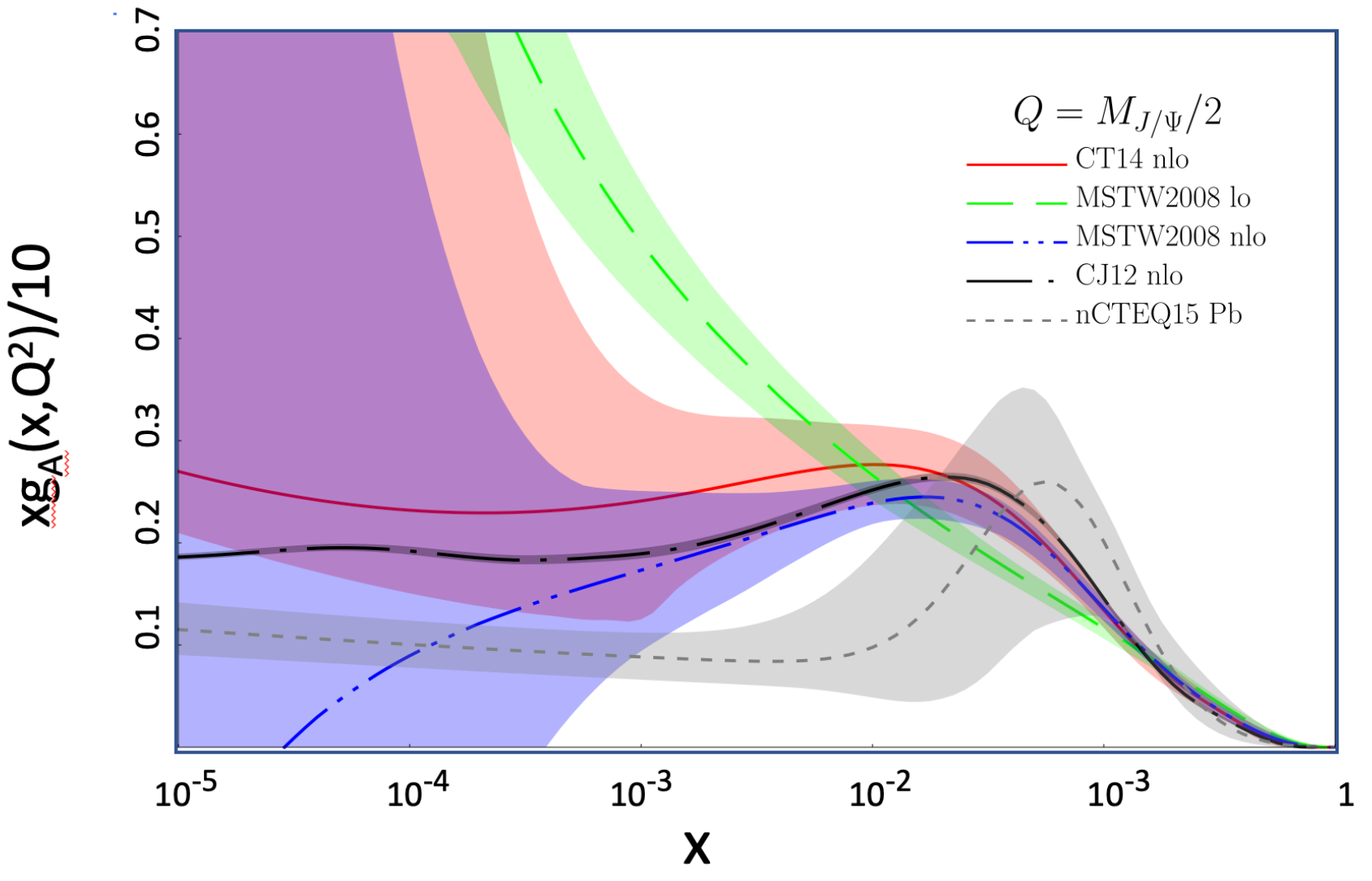}
\caption{\label{pdf}Uncertainties in theoretical compilations of gluon distribution functions as a function of the momentum fraction $x$.}
\end{minipage}%
\end{figure}

In our previous works \cite{Adeluyi:2011rt,Adeluyi:2012ph}, we investigated the influence of various gluon distributions on the production of $J/\psi$ and $\Upsilon(1s)$ vector mesons in UPCs. Notably, UPCs involving pPb and PbPb collisions exhibit distinct production mechanisms termed ``direct" and ``resolved". Direct production entails the photon interacting directly with the nucleus, whereas the resolved mechanism entails the photon fluctuating into a quark-antiquark pair, which subsequently interact with the nucleus. At the leading order, direct production hinges on gluon distributions, that are particularly uncertain within the nucleus, especially at low $x$ values (Figure \ref{pdf}). Conversely, the resolved mechanism probes the distributions of gluons and light quarks in both the photon and nucleus. The quadratic dependency of vector meson production in UPCs heightens sensitivity to gluon distributions in both cross-sections and rapidity distributions \cite{Goncalves:2001vs,Adeluyi:2011rt,Adeluyi:2012ph}.

Our first computations for $J/\Psi$ production, incorporating gluon distributions that account for nuclear gluon shadowing \cite{Adeluyi:2011rt,Adeluyi:2012ph}, align well with experimental data from the ALICE~\cite{Tap13,Abe13,Abe13b,Gru14} collaborations,  as illustrated in Figure \ref{pdf1}. It's evident that $J/\Psi$ and $\Upsilon$ photoproduction in UPCs serves as a potent means to investigate nuclear gluon shadowing in the $x<10^{-3}$ region. 

In contrast to the claims of Refs.~ \cite{Adeluyi:2011rt,Adeluyi:2012ph}, the currently most accepted hypothesis it that the resolved components of the photon are the most important for the production of vector mesons in UPCs. As mentioned in other works~\cite{Guz13,Guz14,Guz16,Guz17,Gonc19,Gonc21,Kov23} many calculations illustrate the important role of hadronic fluctuations of the photon  in the photoproduction of light vector mesons and the large leading twist nuclear gluon shadowing in photoproduction of $q\bar{q}$ pairs off nuclei.

 \section{A History of the Future}

I have illustrated how seemingly straightforward concepts, originating in the 1980s and 1990s, have blossomed into a fertile research domain at relativistic colliders. Experimentally verified unexpected phenomena induced in UPCs encompass a spectrum of discoveries, including double giant resonances, pioneering investigations with anti-atoms,  beam loss attributed to capture of  electrons, light-by-light scattering, exploration of PDFs, and quests for physics transcending the standard model. Could phenomena like $\gamma\gamma \rightarrow {\rm graviton}$ \cite{ANP20} or axion-like particles \cite{GS20,GMR21} be uncovered through this avenue? The verdict remains uncertain, contingent upon further strides in accelerator technology, beam physics, and novel detection methodologies.

But predicting is hard, especially if it is about the future! In my travels through numerous physics departments worldwide, I have often encountered the assertion that nuclear physics lacks the fundamental nature attributed to particle physics. Particle physicists often define ``fundamental" as pertaining to answers regarding interactions, particles, and fields -- like the Higgs -- that bridge the gaps in theories concerning matter and forces in the Universe. To these critics, nuclear physics appears more akin to engineering with nucleons. However, it's evident that questions concerning the  prediction of the Hoyle state in $^{12}$C or the origin of elements  and its implications for the existence of life  must also be regarded as fundamental.

It seems plausible that if supersymmetric particles remain elusive at CERN in the next decade, particle physicists may need to focus on precision calculations and measurements or explore less ``fundamental" avenues, such as the existence of exotic mesons or how medium modifications can modify parton distributions. Traditional particle physicists might feel ashamed. But, this could lead to a diversification of scientific endeavors in laboratory settings.

The future of big science may lie in what some still regard as ``small science." The light source laboratories exemplify this trend, offering the potential to address some of the most pressing questions in nuclear and particle physics. Light, both on-shell and off-shell, proves to be an invaluable tool. It may be time for particle physicists to change their focus towards more practical pursuits and incorporate light-based research into their endeavors.  

In my humble opinion, nuclear physics poses challenges at a significantly higher level compared to other physical sciences. This assertion stems from several factors: (a) The lack of comprehensive understanding of nucleon interaction, (b) the composite nature of nucleons, and (c) the complexity of the nuclear many-body problem arising from various aspects of the strong interaction. Perhaps due to the formidable nature of nuclear physics, senior nuclear physicists often exhibit skepticism and lack of support towards newcomers and new research endeavors.

Alan Bromley expressed in an essay that nuclear physicists are among the harshest referees, often rejecting meritorious papers out of jealousy or excessive criticism and destructiveness~\cite{Brom01}. Conversely, colleagues in other physical disciplines tend to be more supportive of one another. This attitude must evolve for the survival of their own legacy, the nurturing of a new generation of young and talented nuclear physicists, and for the advancement of a science crucial for understanding the universe we inhabit.

\section*{Acknowledgments}

I would like to thank Daniel Tapia-Takaki for beneficial discussions and also for his leadership in this field. This work was partially supported by the U.S. DOE grant DE-FG02-08ER41533, the NSF grant no. 2114669 (Accelnet IAN-QCD https://www.iann-qcd.org) and the Helmholtz Research Academy Hesse for FAIR.

\end{document}